\documentstyle[12pt,psfig,rotate]{article}
\begin{document}
\begin{flushright}
Columbia preprint CU--TP--745
\end{flushright}
\vspace*{1cm}

\renewcommand{\thefootnote}{\fnsymbol{footnote}}
\begin{center}
{\Large Shell-like Structures in \\
an expanding Quark--Antiquark Plasma} \\
\vspace*{1cm}
{{\large \bf Carsten Greiner}$^{1}$\footnotemark[2]
{\large\bf and Dirk-Hermann Rischke}$^{2}$\footnotemark[3],\\[6mm]
$^{1}$ Institut f\"ur Theoretische Physik, Universit\"at Giessen,
\\
D-35392 Giessen, Germany, \\[3mm]
$^{2}$ Physics Department, Pupin Physics Laboratories,
\\
Columbia University, NY 10027, USA.} 
\\[1cm]
\footnotetext[2]{e-mail address: greiner@theorie.physik.uni-giessen.de}
\footnotetext[3]{e-mail address: drischke@nt1.phys.columbia.edu}
{\large April 1996}
\\[2cm]
\end{center}
\thispagestyle{empty}
\renewcommand{\thefootnote}{\arabic{footnote}}
\begin{abstract}
The particle density distribution emerging from
the solution of the Vlasov equation for relativistic, 
non-in\-ter\-acting particles with spherically symmetric initial
conditions is shown to exhibit a shell-like structure 
for late fixed times in the center-of-mass 
(CM) frame of the system. A similar phenomenon was recently observed 
employing the test--particle method to solve the Vlasov equation 
for quarks with Nambu--Jona-Lasinio--type interactions,
and was attributed to the attractive forces among the particles.
Contrary to this claim, it is demonstrated here that this 
effect is of purely relativistic origin and is sensitive only to the 
mass of the particles and their initial phase--space distribution.
\end{abstract}
\newpage
\pagenumbering{arabic}
One of the primary goals in present relativistic heavy--ion collision
experiments at the AGS at Brookhaven, the SPS at CERN,
and in future experiments at Brookhaven's Relativistic Heavy Ion Collider 
(RHIC) and CERN's Large Hadron Collider (LHC) is the
temporary formation and subsequent observation of the so-called 
quark--gluon plasma (QGP), the
deconfined, chirally restored phase of strongly interacting matter 
\cite{QM,CS}. In the common picture of chiral symmetry restoration \cite{CS},
at sufficiently high temperatures or densities the massive
and confined quasi-particle excitations, the constituent quarks, become bare, 
undressed quarks with current quark masses being much smaller than the 
constituent quark masses. 

Recently, Abada and Aichelin \cite{Ai95} studied the
dynamical evolution of a collision-free quark--antiquark system by solving 
the relativistic Vlasov equation 
for the one--particle distribution function $f({\bf x},{\bf p},t)$,
\begin{equation}
\frac{\partial f}{\partial t} \, + \,
\frac{{\bf p}\,c^2}{E} \cdot \mbox{\boldmath  $\nabla $}_{x} f \, - \,
\mbox{\boldmath  $\nabla $}_{x} E \cdot 
\mbox{\boldmath  $\nabla $}_{p} f\, = \, 0 \, \, \, ,
\label{veq}
\end{equation}
where the effective momentum--dependent force 
\begin{equation}
- \mbox{\boldmath  $\nabla $}_{x} E \, = \,
- c~ \mbox{\boldmath  $\nabla $}_{x} [{\bf p}^2+ M_c^2({\bf x},t)\,c^2]^{1/2} 
\, = \, - \frac{M_c({\bf x},t)\, c^2}{E}
~c^2~ \mbox{\boldmath  $\nabla $}_{x} M_c({\bf x},t) \label{pot}
\end{equation}
is generated by a self-consistent gap equation for the constituent
quark mass $M_c$ derived within the chiral quark model of Nambu and 
Jona-Lasinio \cite{NJL,Wi92}. 
The initial condition was 
a sphere of radius $R$, homogeneously filled with quarks and antiquarks 
having a thermal momentum distribution corresponding to a temperature $T$
well above the transition temperature $T_c \sim 140$ MeV.
Such ``hot spots'', i.e., small--scale fluctuations with
sufficiently high (and thermalized) energy density, 
are expected to occur in present
heavy--ion experiments at the SPS \cite{We94} and also in future collider 
experiments at RHIC \cite{Ri96}.

The striking observation of Abada and Aichelin was that, in the evolution
of the system, most particles
remain concentrated in an expanding shell, and that
consequently the phase transition
takes place at the inner and outer surface
of that shell. Thus, the situation is fundamentally different from the
naive expectation that the initial fireball expands by retaining its
shape and consequently hadronizes from its surface only.
For the interpretation of the results, Abada and Aichelin argued as follows:
in the non-relativistic case the thermal velocity distribution of particles
peaks around $\langle v \rangle = \sqrt{2T/m}$, and
in an interaction--free expansion, all the many particles with that velocity
will stay together, leading to a corresponding peak in
the density, i.e., the observed shell-like structure. 
For interacting quarks, the attractive forces in 
the Nambu--Jona-Lasinio model will enhance this effect by trying to 
keep the (bare) quarks in the region of high density.
This conclusion was further motivated by a similar
observation found in non-relativistic simulations
of expanding nuclear matter, where a shell structure forms
after the nucleons have frozen out and enter the liquid--gas transition regime
\cite{Gr92}.

In contrast, in this note we show that the observation by Abada and Aichelin
\cite{Ai95}, i.e., the  shell-like structure in the density distribution,
originates simply from ultrarelativistic kinematics.
Moreover, contrary to their line of arguments, we show that in the
non-relativistic limit, the density distribution does {\em not} exhibit
such a structure.
We will not justify the validity of the Vlasov equation 
for modelling the expansion of a quark--antiquark plasma. Obviously,
collisions among the constituents are important to achieve the assumed,
thermally equilibrated initial condition in the first place. 
The relevant quantity for thermalization is given by the transport rate
$\Gamma_q^{trans}$ of the quarks in a QGP \cite{Th94}.
Estimates with a simple infrared cutoff suggest a rather high
rate of $\Gamma_q^{trans} \approx 0.3 - 0.5\, T$ 
($T$ denotes the temperature of the 
plasma), however, a more detailed treatment of the infrared sector
in those rates by means of resummation methods raises the question whether
one can trust those perturbative calculations at all \cite{Th94}.

The collision rate may be high enough to maintain (local) thermal
equilibrium even during the expansion. In that case
the system's evolution is governed by ideal hydrodynamics.
We shall compare the relativistic
hydrodynamical solution for an expanding sphere 
with the results obtained from solving the Vlasov equation.

On the other hand, it was also recently speculated that, in particular 
shortly before the phase transition happens,
the collisions in the plasma become so rare that their effect
on the overall dynamics can be neglected. 
Then, the system freezes out already before the transition occurs
and it is appropriate to model the non-thermal dynamics 
with the Vlasov equation \cite{Cs95}.

There is a very simple, intuitive explanation why a shell occurs
in the relativistic expansion of the fireball:
expanding matter piles up in a shell, because the velocity of light
$c$ limits the maximum particle velocity.
Suppose the particles are ultrarelativistic, i.e., massless.
In that case, the velocity of {\em all} particles equals the velocity of light,
$v \equiv c$. Obviously, without collisions, 
at times $t \gg R/c$ they can only be located in
a spherical shell of thickness $2\, R$, moving outward with
light velocity. (The particle density in that shell is not necessarily 
constant, but depends on the initial distribution of momenta and coordinates 
in the sphere, see below).
The interior of the shell will be completely void of particles.
For particles with finite mass,
the thickness of the shell is roughly
$$
d \, \approx \, 2R + \Delta v \, t \, \, \, ,
$$
where $\Delta v \equiv |\langle v^2 \rangle - \langle v \rangle^2|^{1/2}$ 
characterizes the average spread of the velocity distribution. 
That spread grows with increasing mass of
the particles, such that in the non-relativistic limit the interior
of the shell is filled with particles, thus restoring the naive
expectation for the expansion of a sphere (and disproving the
arguments presented in \cite{Ai95}).

As will be argued below,
the modification of the momenta and velocities of the particles due to
the force term (\ref{pot}) and
the time dependence of the mass are small,
so that the generic features of phase--space dynamics will
be represented by the free--streaming limit of eq.\ (\ref{veq}),
\begin{equation}
\frac{\partial f}{\partial t} \, + \,
{\bf v} \cdot \mbox{\boldmath  $\nabla $}_{x} f \, = \, 0 \, \, \, ,
\label{fveq}
\end{equation}
where ${\bf v} \equiv {\bf p}c^2/E$ and the mass $M_c$ of the quarks
is assumed to be independent of time.
In the following we explicitly construct the solution for the particle
density distribution\footnote{We choose units $\hbar=k_B=1$.}
\begin{equation}
\rho ({\bf x},t) \, = \, \int \frac{d^3{\bf p}}{(2\pi )^3} \, f({\bf x},
{\bf p},t)~.
\label{rho}
\end{equation}
Our initial condition will ultimately be the same as in \cite{Ai95}, i.e.,
the particles are thermally distributed in momentum space 
and homogeneously distributed in coordinate space
within a sphere of radius $R$. For the moment, however, arbitrary 
(though homogeneously filled) initial coordinate--space
volumina $V$ and arbitrary initial momentum distributions $n({\bf p})$ are
allowed.

We introduce the characteristic function of the initial volume,
$\Theta_V ({\bf x}) = 1$ for ${\bf x} \in V$, and 0 elsewhere.
Then, for a coordinate--space distribution which is homogeneous in $V$,
the initial phase--space distribution separates as
\begin{eqnarray}
f({\bf x},{\bf p},0) & = & \Theta_V ({\bf x}) ~ n({\bf p})
\nonumber \\
&\equiv & \int d^3{\bf a} \,\, \Theta_V({\bf a}) \,
f_{{\bf a}}({\bf x},{\bf p},0)
\label{fact}
\end{eqnarray}
where we defined the phase--space distribution for a $\delta$--like
source,
\begin{equation}
f_{{\bf a}}({\bf x},{\bf p},0) \, = \,
\delta ^3({\bf x} - {\bf a})\,  n({\bf p}) \, \, \, .
\label{fa}
\end{equation}
Since (\ref{fveq}) is a homogeneous linear differential equation,
the solution $f({\bf x}, {\bf p}, t)$ has the form
\begin{equation}
f({\bf x}, {\bf p},t) \, = \,
\int d^3{\bf a} \,\, \Theta_V ({\bf a}) \,
f_{{\bf a}}({\bf x},{\bf p},t) \, \, \, .
\label{solf}
\end{equation}
The solutions $f_{{\bf a}}$ to eq.\ (\ref{fveq})
(for time--independent particle mass) are readily obtained:
\begin{equation}
f_{{\bf a}}({\bf x},{\bf p},t) \, = \,
f_{{\bf a}}({\bf x}-{\bf v}t,{\bf p},0) \, = \,
\delta ^3({\bf x} -{\bf v}t - {\bf a}) 
\, n({\bf p}) \, \, \, .
\label{solfa}
\end{equation}
The particle density distribution (\ref{rho}) $\rho _{{\bf a}}({\bf x},t)$ 
for a $\delta$--like source (\ref{fa}) located at ${\bf x}={\bf a}$ 
follows after a careful evaluation of the $\delta $-function as
\begin{eqnarray}
\rho _{{\bf a}}({\bf x},t) & = &
\int \frac{d^3{\bf p}}{(2\pi )^3} \, n({\bf p}) \,
\delta ^3\left({\bf x} -\frac{{\bf p}\, c^2}{E}t - {\bf a}\right)
\nonumber \\
&=&
\frac{1}{(2\pi )^3} \,
n(m \cosh \eta \, {\bf u})
\, \cosh^5 \eta \left( \frac{m}{t} \right)^3
\, \Theta \left(1-\tanh \eta \right) \, \, \, , \, \, t>0\, ,
\label{rhoa}
\end{eqnarray}
where
\begin{equation}
{\bf u} \equiv \frac{{\bf x}-{\bf a}}{t}~, \, \, \,  
\tanh \eta \equiv |{\bf u}|/c~.
\end{equation}
Obviously, $m \cosh \eta \,{\bf u}$ is the momentum and
$mc^2 \cosh \eta$ the energy of the particles.
The step function assures causal expansion.
The non-relativistic limit corresponds to
$|{\bf u}| \ll c,\, \cosh \eta \approx 1$ and (\ref{rhoa}) becomes
\begin{equation}
\rho ^{{\rm n.r.}}_{{\bf a}}({\bf x},t) \,  = \,
\frac{1}{(2\pi )^3}  \,
n \left( m \,{\bf u} \right) \,
\left( \frac{m}{t} \right) ^3
\, \, \, .
\label{rhoanrel}
\end{equation}
If the initial momentum distribution $n({\bf p})$ is isotropic, i.e.,
$n(|{\bf p}|)$, and a monotonously decreasing function of $|{\bf p}|$, 
as it is the case for a thermal (Bose--Einstein, Fermi--Dirac, or 
Maxwell--Boltzmann) distribution, the
density distribution $\rho_{{\bf a}} ({\bf x},t)$ is also isotropic, 
$\rho_{\bf a} (r,t)$, and, 
in the non-relativistic limit, a decreasing function of the (radial)
distance $r \equiv |{\bf x} - {\bf a}|$ from the 
source. This disproves the argument presented in \cite{Ai95} about
the formation of a shell-like structure in the non-relativistic limit.

This behaviour changes completely in the ultrarelativistic 
regime due to the additional factor $\cosh^5 \eta$. This term is a
monotonously increasing function of $r$
and, in combination with the momentum distribution function, thus 
leads to a pronounced peak in the density distribution.
In Fig.\ 1 the quantity $(ct)^3\, \rho _{\bf a} (r,t)$ 
is plotted as a function of $r/ct$ for various
masses $m$. (Note that this representation is time--invariant.)
A Fermi distribution
$$
n (|{\bf p}|) \, \equiv \, n_F(|{\bf p}|) \, = 
\, \frac{1}{e^{E/T}+1}~,~~E \equiv c\, \sqrt{{\bf p}^2 
+(mc)^2}~,
$$
was chosen with a temperature $T=160$ MeV, slightly larger
than the expected critical temperature $T_c$.
The shape of the distribution is
very sensitive to the ratio $m c^2/T$.
One clearly sees how the peak structure emerges already when
$m c^2/T$ is of order one, as most of the particles are already
moving with a velocity close to $c$.
(If $m c^2/T \ll 1$ the width of the velocity distribution is proportional to
$ \sim (m c^2/T)^2$.)

For a uniformly filled sphere of radius $R$,
$$
\Theta _V ({\bf a}) \, = \left\{
\begin{array}{lc}
1 & \mbox{for} \, |{\bf a}| \leq R \\
0 & \mbox{for} \, |{\bf a}| > R
\end{array}
\right.
\, \, \, .
$$
The particle density distribution $\rho ({\bf x},t)$ is then obtained as
\begin{equation}
\rho ({\bf x},t) \, =  \int d^3{\bf a} \,\,
\Theta _V({\bf a}) \,
\rho _{{\bf a}} ({\bf x},t) \, \, \, .
\label{dens}
\end{equation}
In principle, one should multiply (\ref{dens}) by the degeneracy 
$d_q =12$ of light quarks (and similarly for the antiquarks), but this factor 
is, for
the sake of simplicity, omitted in the following. Eq.\ (\ref{dens})
can be reduced to a one-dimensional integral and takes the form
\begin{eqnarray}
\rho (\tilde{r},\tilde{t}) & = &
\left( \frac{mc}{2\pi \tilde{t}} \right)^3 \frac{\pi}{\tilde{r}} 
\, \Theta (\tilde{t} - \tilde {r} +1) \,
\int _{\tilde{r}-1}^{{\rm min}(\tilde{r}+1;\tilde{t})}
dz \, z \left( 1 - (z-\tilde{r})^2\right) \, n(mc \sinh \eta_z) \, 
\cosh^5 \eta_z 
\nonumber \\[5mm]
&& \mbox{for} \, \, \tilde{r} \, > \, 1 \, \, \, ;
\nonumber \\
&&
\label{denssol}
\\
\rho (\tilde{r},\tilde{t}) & = &
\left(\frac{mc}{2\pi \tilde{t} } \right)^3 \, 2 \pi \left[ \,
2 \int _{0}^{{\rm min}(1-\tilde{r};\tilde{t})}
dz  \,  z^2 \, n(mc \sinh \eta_z) \, \cosh^5 \eta_z
\right. \nonumber \\
&& \hspace*{15mm} \left.
+ \, \frac{1}{2\tilde{r}}\, \Theta (\tilde{t} + \tilde {r} -1 ) \,
\int _{1-\tilde{r}}^{{\rm min}(1+\tilde{r};\tilde{t})}
dz\, z \left( 1 - (z-\tilde{r})^2\right) \, n(mc \sinh \eta_z) \, 
\cosh^5 \eta_z
\, \right]
\nonumber \\[5mm]
&& \mbox{for} \, \, \tilde{r} \, < \, 1 \, \, \, ,
\nonumber
\end{eqnarray}
where the dimensionless variables $\tilde{r}=r/R$ and $\tilde{t}=ct/R$ have
been introduced and $\tanh \eta_z \, \equiv \, z/\tilde{t}$.
Expression (\ref{denssol}) has to be integrated numerically.

In Fig.\ 2 the resulting density profile $\rho (\tilde{r},\tilde{t})$ is
shown for different times $\tilde{t}$ between 0.1 and 3. Again, the initial
momentum distribution is a Fermi distribution 
$n_F$ with a temperature $T=160$ MeV.
The mass of the quarks is taken as
$m (\equiv M_c)=50 \,$MeV$/c^2$. For times $\tilde{t}<1$ the density in
the central region $\tilde{r} \leq 1 - \tilde{t}$ remains constant
due to causality. In other words, it remains constant because
as many particles leave an infinitesimal volume element as do enter.
However, since most particles move with nearly light velocity
(the thermal velocity is $\langle v \rangle \approx 0.9858 \, c$
for the present situation, while the spread of the velocity
distribution is only $\Delta v \equiv
|\langle v^2\rangle - \langle v \rangle^2|^{1/2} \approx 0.0031\, c$), 
for times $\tilde{t}>1$ no particles can enter the central region
around $\tilde{r} =0$ any more, so that it becomes completely depleted. 
Thus, a shell starts to move outwards 
at time $\tilde{t}=1$, as one clearly recognizes in Fig.\ 2.
As expected, that shell has a width of $2\, R$ and travels
with the velocity of light.

In Fig.\ 3  density profiles $\rho (\tilde{r},1.5)$
are shown for different masses. For all masses smaller than 300 MeV$/c^2$ 
relativistic effects become important, and a shell structure emerges in
the temporal evolution.

This behaviour is quite similar to the relativistic, ideal hydrodynamical 
expansion of a spherical fireball \cite{Bi95}. Also in this case, the particle
density distribution exhibits a shell structure at fixed times in the CM
frame of the fireball. The reason is that moving matter at finite $r$
experiences relativistic time dilation \cite{Ri95} with respect to
matter in the center (which is at rest due to symmetry) and thus
dilutes less rapidly. To confirm this, and to investigate the effect
of a varying particle mass on the expansion, 
we solved the equations of ideal relativistic hydrodynamics,
\begin{equation} \label{hydro}
\partial_{\mu} T^{\mu \nu} = 0 
\end{equation}
($T^{\mu \nu}=(\epsilon+p)u^{\mu}u^{\nu}-p g^{\mu \nu}$ 
is the energy--momentum tensor of an ideal fluid, $\epsilon,\, p$ are
energy density and pressure in the local rest frame of a fluid element,
moving with 4--velocity $u^{\mu}$ in the CM frame of the fireball, 
$g^{\mu \nu} = {\rm diag}(+,-,-,-)$ is the metric tensor), for
a spherically symmetric initial fireball of radius $R$ and temperature
$T=160$ MeV and for ideal Fermi gas equations of state
\begin{eqnarray}
\epsilon (T) & = & \frac{1}{2\pi^2 } \int_0^{\infty} {\rm d}p\, p^2
\sqrt{p^2 + m^2c^2} \,\, n_F(p)~, \\
p(T) & = & \frac{1}{2\pi^2} \int_0^{\infty} {\rm d}p\, \frac{p^4}{
\sqrt{p^2 + m^2c^2}} \,\, n_F(p)~,
\end{eqnarray}
with masses $m$ varying between 10 and 940 MeV/$c^2$.
The corresponding particle densities are
\begin{equation}
\rho(T) = \frac{1}{2\pi^2} \int_0^{\infty} {\rm d}p\, p^2 \,\, n_F(p)~.
\end{equation}
The hydrodynamical equations (\ref{hydro}) are solved with the
(one--dimensional) relativistic Harten--Lax--van Leer--Einfeldt algorithm
\cite{Ri95}, supplemented
with a Sod predictor--corrector step \cite{Sod} to account for spherical
geometry \cite{Ri96b}.

In Fig.\ 4 we show the corresponding
density profiles at time $\tilde{t}= 2.97$. The situation is qualitatively
very similar to Fig.\ 3, which is remarkable considering the fact
that ideal hydrodynamics corresponds to the limit of an infinite collision
rate (and thus immediate local thermodynamical equilibrium). 
The main quantitative differences are
(a) the time $\tilde{t}$ has to be about twice as large as before to
obtain density distributions similar to Fig.\ 3, and (b)
the quark densities are smaller.
The reason for (a) is that for a time $\tilde{t}=1.5$ the hydrodynamical 
rarefaction has not yet reached the center of the fireball (the rarefaction
velocity is the velocity of sound $c_s \leq c/\sqrt{3} \approx 0.5774\, c$,
in contrast to the free--streaming case where the rarefaction wave travels
with the velocity of light $c$),
and thus the shell structure could not possibly have formed. 
The reason for (b) is that the quark density is not conserved in the 
solution of (\ref{hydro}), in contrast to the conservation of the quark number
in the free--streaming scenario: due to the work performed in the
hydrodynamical expansion the system cools, and thus
temperature and, consequently, particle density decrease.
Thus, the densities are smaller in Fig.\ 4 than in Fig.\ 3. 
In order to conserve the quark number in the hydrodynamical evolution, 
one would have to supplement
the equations of motion (\ref{hydro}) with the continuity equation for
the quark number current, but this would also require introducing a
finite quark chemical potential in the equation of state. This is
certainly beyond the scope of our present work, where we are interested
in qualitative similarities between relativistic free--streaming and
hydrodynamical solutions.

Finally, we address the potential influence of the
self-consistent mass term and the corresponding force (\ref{pot})
on the expansion dynamics. The mass $M_c(r,t)$ influences
the velocities of the particles in two ways.
First, due to its spatial dependence a force is generated which changes
the momentum of a particle in time according to
\begin{equation}
\dot{{\bf p}} \, = \, \dot{p} \, {\bf e}_{{\bf p}} \, + \, p \, \dot{{\bf 
e}}_{{\bf p}}
\, = \, -\frac{M_c(r,t)c^2}{E}\, c^2 \,
\mbox{\boldmath  $\nabla $}_{x}M_c(r,t)
\, \, \, .
\label{force}
\end{equation}
Second, the velocity of a particle changes because of the
time dependence of the mass; in particular it will continuously decrease
if $M_c(t)$ increases with time as in the present case.

According to \cite{Ai95} the initial mass of the particles is $M_c \approx
50 $ MeV/$c^2$. For a temperature of $T\approx 240$ MeV (used in \cite{Ai95})
the thermal velocity is $\sim 0.9931 \, c$.
To estimate the typical change in momentum, the modulus of 
the gradient in the mass $M_c$ is (conservatively)
read off the results of \cite{Ai95} as
$|\mbox{\boldmath  $\nabla $}_{x}M_c(r,t)\, c^2| \leq 500$ MeV/fm,
acting over a period in time $\Delta t \approx 0.5$ fm/$c$.
Taking $\langle E \rangle$ and $\langle p\, c \rangle
\,  \approx \, 3T$ and an average mass
$\langle M_c\, c^2\rangle \approx 100$ MeV the change in the (average)
momentum is estimated as
\begin{eqnarray}
\frac{\langle \Delta p \rangle}{ \langle p \rangle} & = &
\frac{\langle \dot{p}\rangle  \Delta t}{\langle p \rangle} \, \leq \,
\frac{1}{\langle p \rangle} \, \frac{\langle M_c\, c^2\rangle }{\langle E 
\rangle}\, |\mbox{\boldmath  $\nabla $}_{x}M_c(r,t)c^2|\,
\Delta t \, \approx \, 0.05
\label{estimate}
\\[2mm]
\langle \Delta {\bf e}_{{\bf p}} \rangle & =  &
\langle \dot{{\bf e}}_{{\bf p}} \rangle \Delta t \, \leq \,
\,  \frac{1}{\langle p \rangle} \, \frac{\langle M_cc^2\rangle}{\langle
E \rangle}\, |\mbox{\boldmath  $\nabla $}_{x}M_c(r,t)c^2|\,
 \Delta t \, \approx \, 0.05 \, \, \, .
\nonumber
\end{eqnarray}
Hence, the typical momentum of a particle is to good approximation conserved
both in magnitude and direction; the change is at most a few percent.

For approximate momentum conservation, $p_i(t=t_i)c \approx 3T \approx
p_f(t=t_f)c$,
the change in the typical velocity of the particles due to the increase of the 
mass
from $M_c(t_i)\approx 50$ MeV$/c^2$ to
$M_c(t_f)\approx 300$ MeV$/c^2$ is readily obtained as
\begin{equation}
v _{i;M_c =50\, {\rm MeV}/c^2} -
v_{f;M_c =300\, {\rm MeV}/c^2}  \, \approx \,  0.09 \, c \, \, \, .
\label{deltav}
\end{equation}
All particles with large (thermal) velocities will be decelerated
by typically ten percent. As a consequence, the formation of the shell
will be somewhat delayed. Furthermore, this deceleration of the
particles will stabilize the region of
expanding matter, where the mass of the particles is still rather low, i.e.,
the interior of the outgoing shell.
In conclusion, taking a self-consistent mass into account will only have
minor effects on the dynamics. The generic structure of phase--space
dynamics is dominated by relativistic kinematics as given by the solution
of (\ref{fveq}).

To conclude, in this note we have investigated 
the free--streaming expansion of a fireball consisting of quarks and/or
antiquarks. We explicitly constructed the solution for spherically
symmetric initial conditions and confirmed the existence 
of a shell-like structure in the expansion at late times \cite{Ai95}.
Contrary to the arguments of \cite{Ai95}, we have
explained that structure as arising solely from (and only for)
relativistic kinematics. Interestingly enough, similar structures occur
in the ideal hydrodynamical limit (where interaction rates are infinite), 
which again confirms that their origin is kinematical rather than due to
specific details of the interaction between the constituent particles.

The overall decrease in the density of the shell is $\sim 1/t^3$, cf.\
eq.\ (\ref{denssol}). 
If the critical density for the
phase transition is reached when the shell was already formed
(for times
$t \geq R/c$, cf.\ Fig.\ 2),
the transition necessarily happens at the outer as well as the inner surface
of the shell. 
This is what has been observed in \cite{Ai95}. In principle, however, it
is also conceivable that the phase transition happens before the shell
structure has formed, for instance when the initial temperature is
only slightly above the critical temperature $T_c$. 
Then the fireball will hadronize only in the surface region where the density
decreases fast at times $t < R/c$.

The analytic solutions presented here may also serve as benchmark tests
for numerical algorithms like the test--particle method. Another 
test case with a solution that can be readily obtained by similar
means (but will not be presented here for the sake of brevity)
is the free--streaming expansion of a slab (finite thickness in one dimension,
but infinite extension in the other two). In that case, no shell structure
emerges (because sufficiently many particles enter the central region
from other parts of the slab). This is in contrast to the analogous
hydrodynamical expansion problem \cite{Ri95}.
\\[15mm]
{\bf Acknowledgements:}
\\[5mm]
Discussions with J.\ Aichelin are acknowledged.
Both authors thank the Alexander von Humboldt Stiftung
for partial support under the Feodor Lynen program.
C.G.\ acknowledges support by the BMBF and GSI Darmstadt, D.H.R.\ 
acknowledges support by the Director, Office of Energy
Research, Division of Nuclear Physics of the Office of 
High Energy and Nuclear Physics of the U.S.\ Department of 
Energy under Contract No.\ DE-FG-02-93ER-40764.
\newpage
\parskip0mm
\par
\newpage
\parindent 0mm
{\Large {\bf Figure captions}}:
\\[2cm]
{\bf Figure 1}:  \\
The density profile $(ct)^3 \rho_{\bf a} (r,t)$
of a point source for various masses $m$ as a function of $r/ct$.
The initial momentum distribution is a Fermi distribution
with temperature $T=160$ MeV.
\\[1cm]
{\bf Figure 2}:  \\
The density profile $\rho (\tilde{r},\tilde{t})$ 
at various stages $\tilde{t}$ of the free--streaming expansion of
a fireball. The initial momentum distribution is a Fermi distribution
with $T=160$ MeV. The mass of the particles is taken as
$m =50$ MeV/$c^2$. At times $\tilde{t}>1$ a shell structure
emerges in the radial expansion.
\\[1cm]
{\bf Figure 3}: \\
The density profile $\rho (\tilde{r},\tilde{t})$ 
at a fixed time $\tilde{t}=1.5$ for the free--streaming expansion of
a fireball. The mass of the particles is varied between
10 and 940 MeV/$c^2$ to demonstrate the relativistic effects in the expansion.
\\[1cm]
{\bf Figure 4}: \\
The density profile $\rho (\tilde{r},\tilde{t})$ 
at a fixed time $\tilde{t}=2.97$ for the
ideal relativistic hydrodynamical expansion of
a fireball. The equation of state is that of an ideal gas of
relativistic particles with mass varying between 10 and 940 MeV/$c^2$.

\begin{figure}[htbp]
\begin{minipage}{10cm}
{\psfig{figure=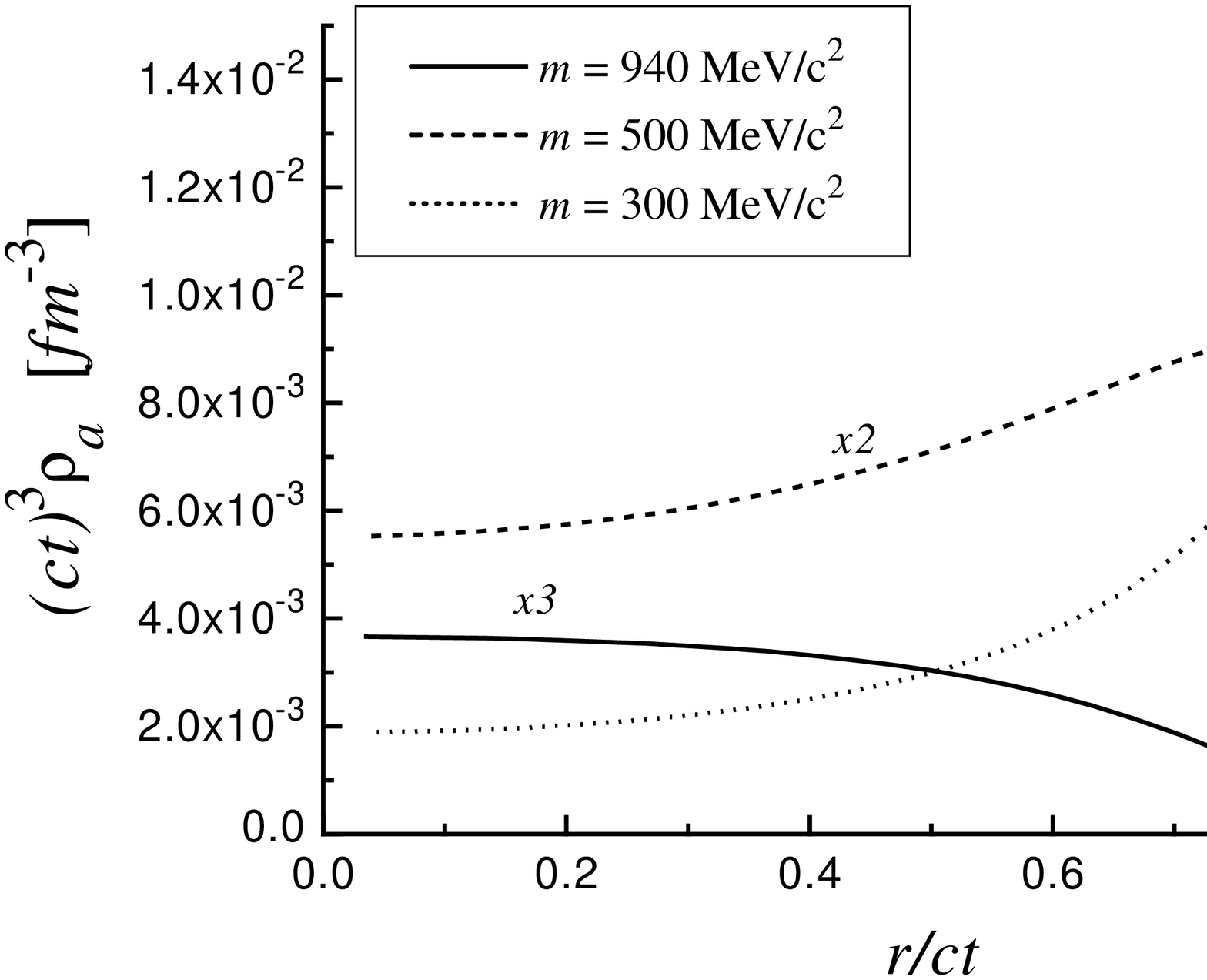,width=10cm}}
\end{minipage}
\vspace{-0.4cm}
\end{figure}
\newpage
\begin{minipage}{10cm}
{\psfig{figure=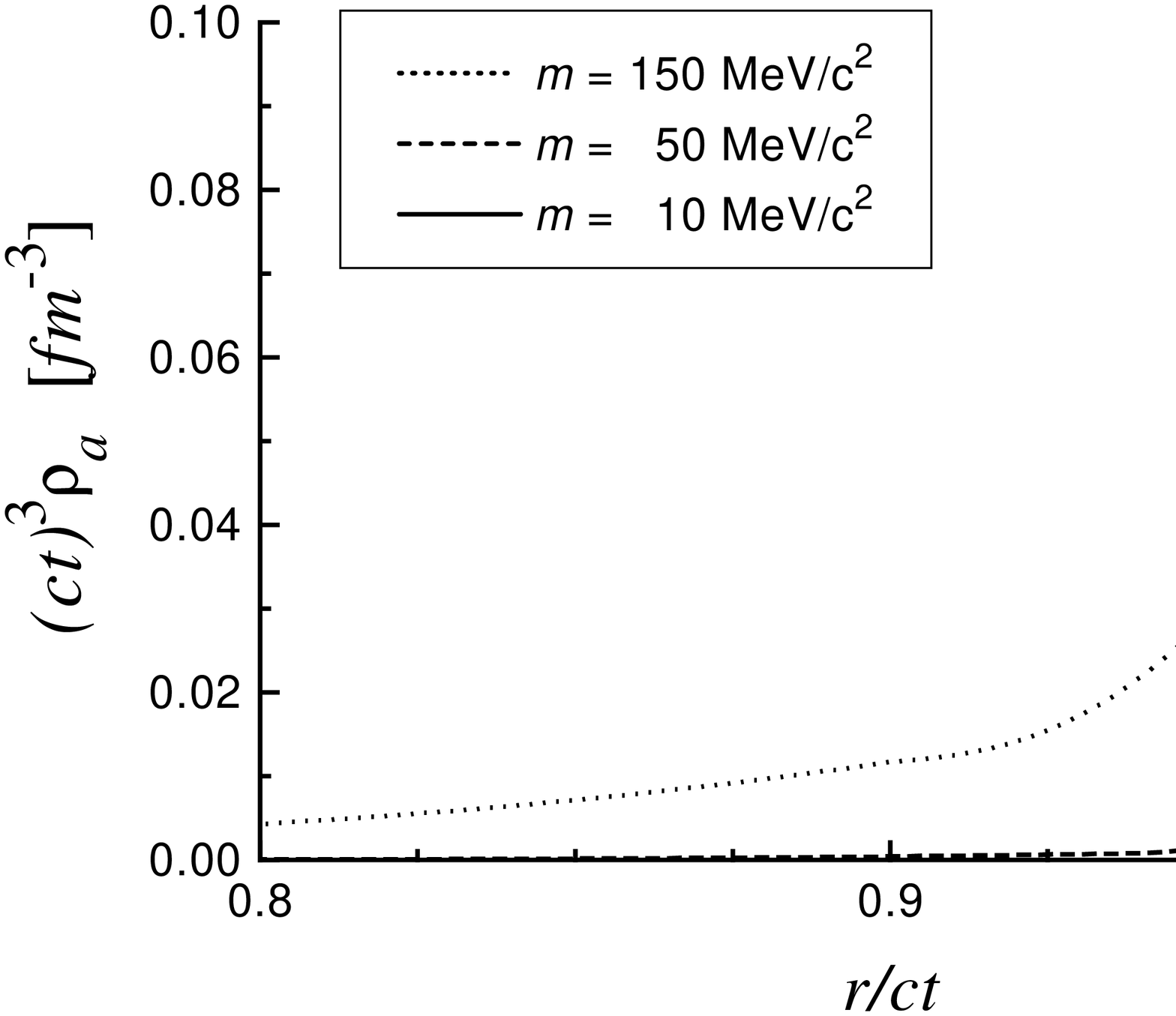,width=10cm}}
\end{minipage}
\vspace{-0.4cm}
\newpage
\begin{minipage}{10cm}
{\psfig{figure=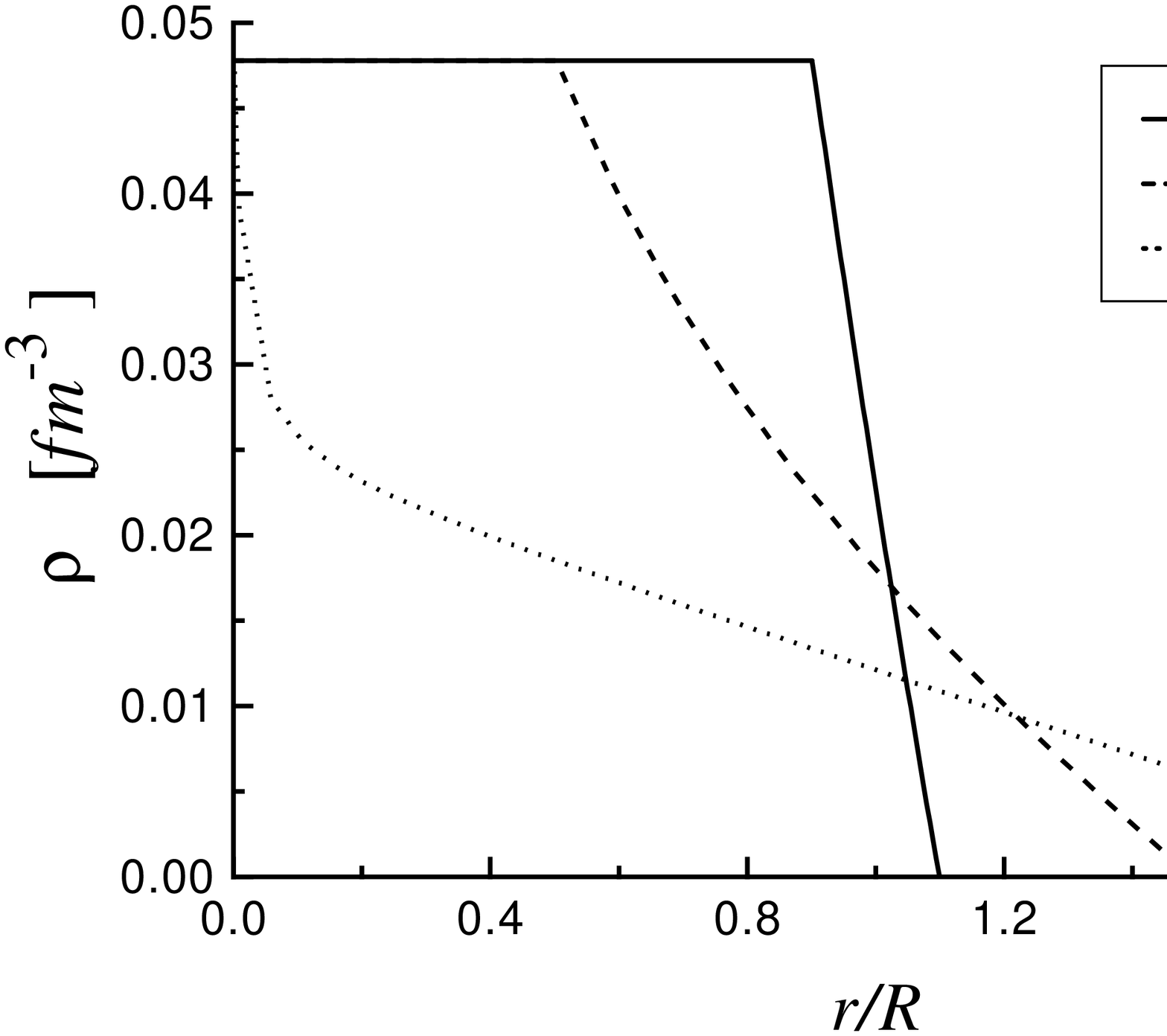,width=10cm}}
\end{minipage}
\vspace{-0.4cm}
\newpage
\begin{minipage}{10cm}
{\psfig{figure=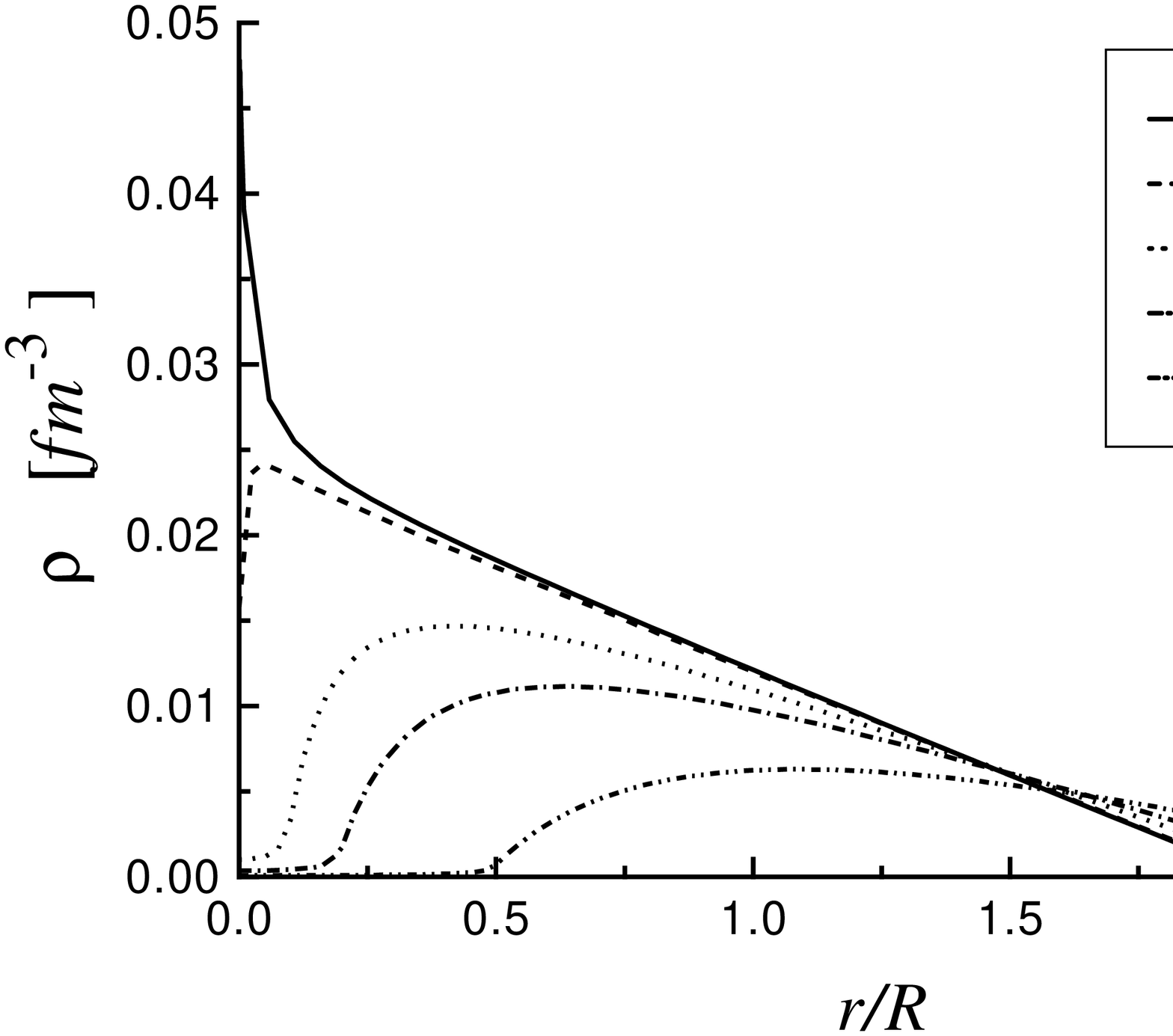,width=10cm}}
\end{minipage}
\vspace{-0.4cm}
\newpage
\begin{minipage}{10cm}
{\psfig{figure=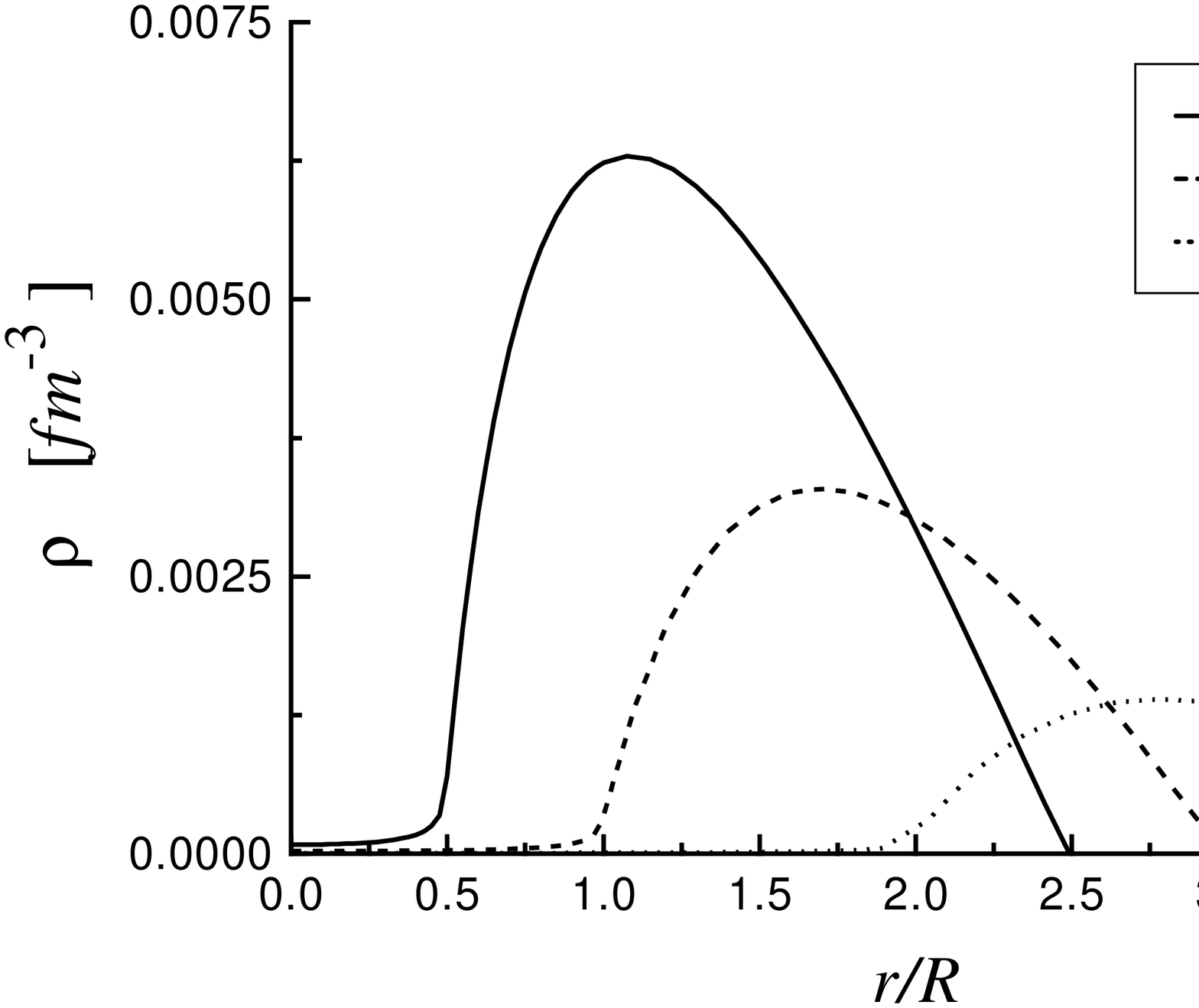,width=10cm}}
\end{minipage}
\vspace{-0.4cm}
\newpage
\begin{minipage}{10cm}
{\psfig{figure=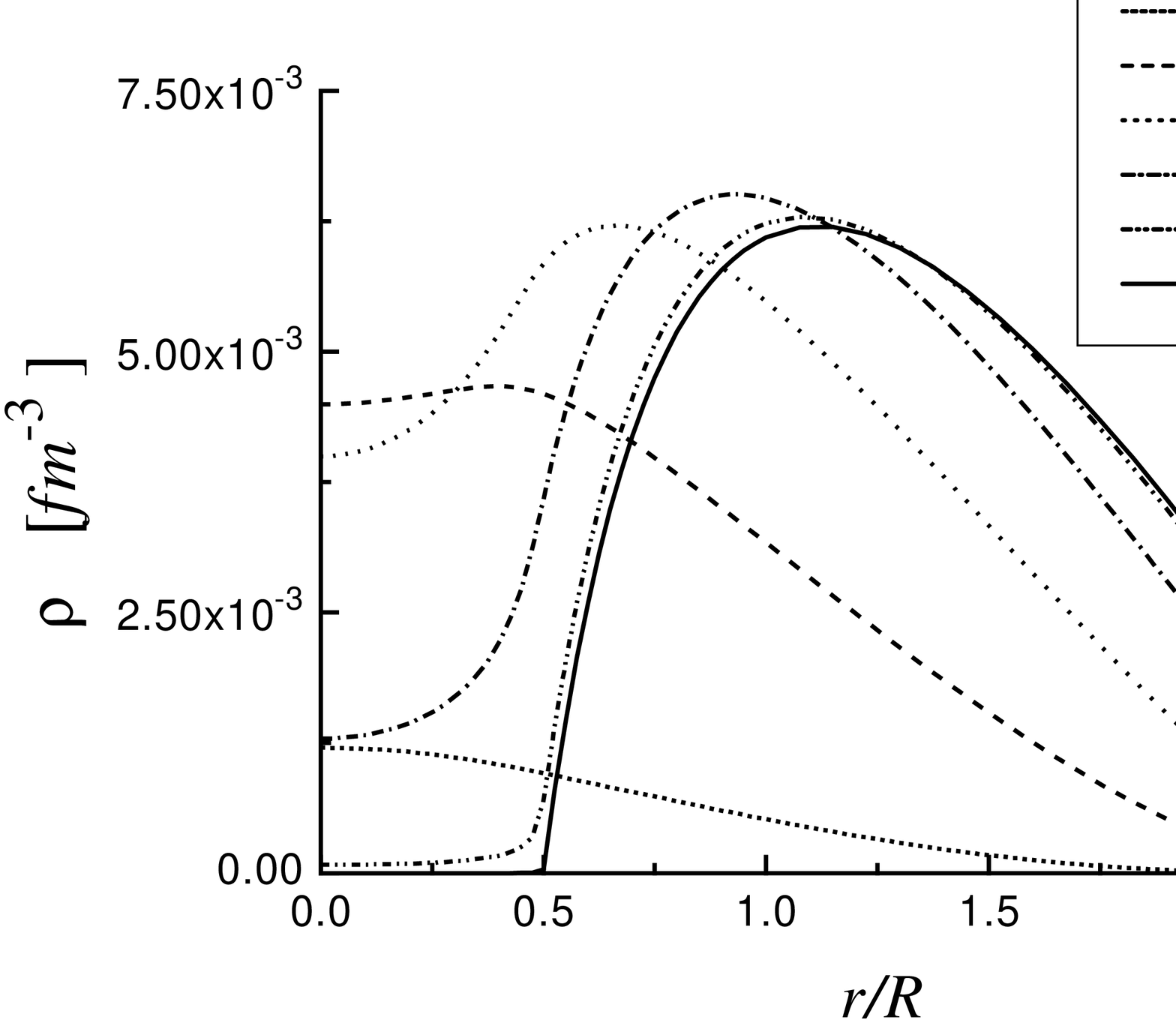,width=10cm}}
\end{minipage}
\vspace{-0.4cm}
\newpage
\begin{center}
\begin{minipage}{10cm}
\rotate[r]{{\psfig{figure=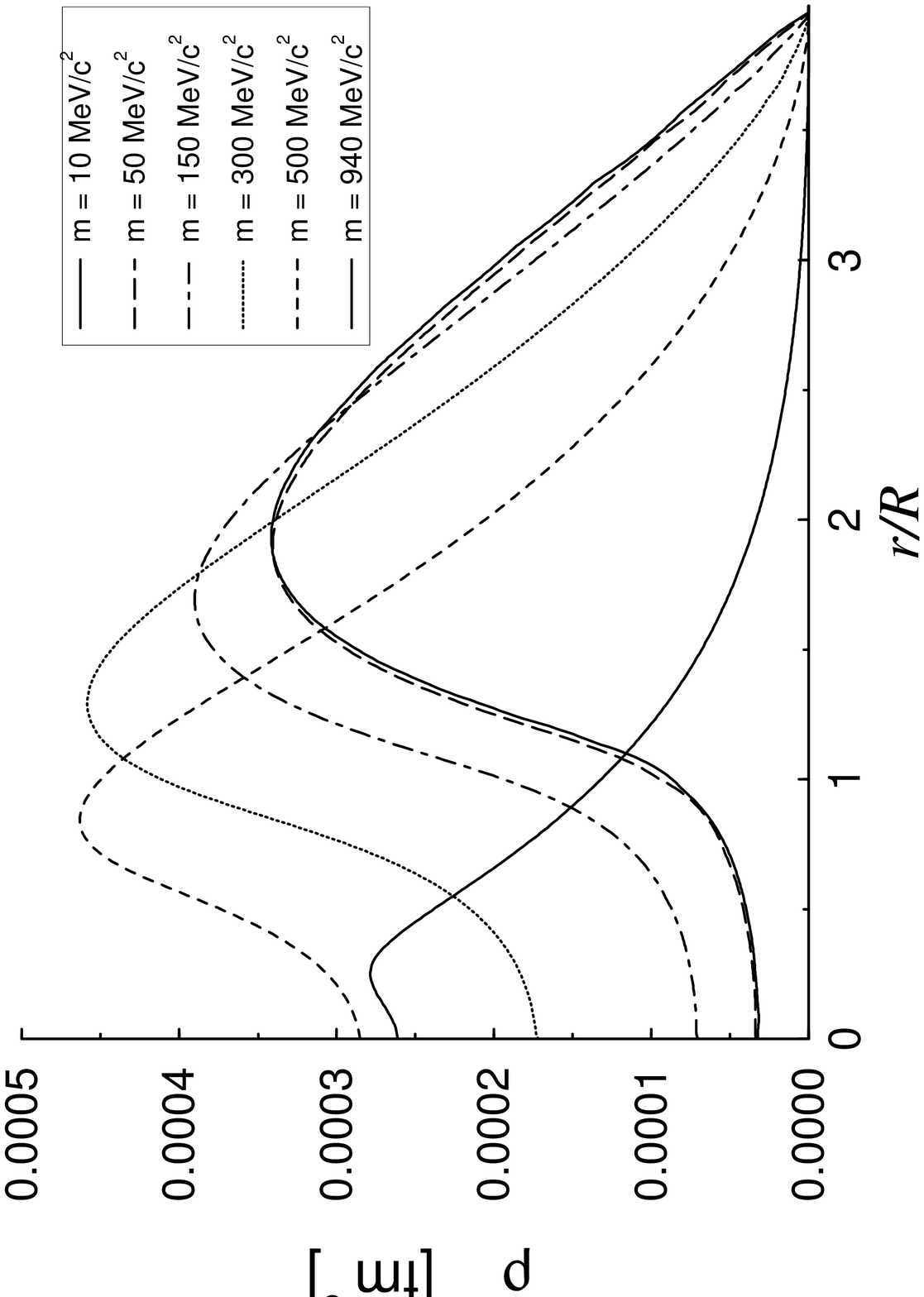,width=10cm}}}
\end{minipage}
\end{center}
%
%
%
%
%
%
\end{document}